\documentstyle[aps,prb]{revtex}
\input{epsf}
\begin{document}
\draft

\title{Deviations from the Gaussian distribution of mesoscopic 
conductance fluctuations}
\author{M. C. W. van Rossum$^1$\footnote{Present address:
Room 123,
Anatomy-Chemistry Building,
University of Pennsylvania,
Philadelphia, PA 19104-6058, USA,
e-mail: vrossum@retina.anatomy.upenn.edu},
Igor V. Lerner$^2$, Boris L. Altshuler$^3$, 
and Th. M. Nieuwenhuizen$^1$} 
\address{$^1$ Van der Waals-Zeeman Laboratorium, Universiteit van Amsterdam
\\ Valckenierstraat 65, 1018 XE Amsterdam, The Netherlands\\
$^2$ School of Physics, University of Birmingham, Edgbaston, Birmingham B15
2TT, UK \\
$^3$ NEC Research Institute, 4 Independence Way, Princeton, NJ 08540, USA\\}
\date{\today}
\maketitle

\begin{abstract} The conductance distribution of metallic
mesoscopic systems is considered. The variance of this distribution describes the universal
conductance fluctuations, yielding a Gaussian distribution of the conductance.
We calculate diagrammatically the third cumulant of this distribution, the
leading deviation from the Gaussian. We confirm random matrix theory
calculations that the leading contribution in quasi-one dimension vanishes.
However, in quasi two dimensions the third cumulant is negative, whereas in
three dimensions it is positive. \end{abstract}


\section{Introduction}\label{intro}

About ten years ago the universal conductance fluctuations (UCF) were
discovered in mesoscopic, metallic samples\cite{umbach,lee,altshuler2}. The
electronic conductance of these samples shows reproducible sample to sample
fluctuations. The fluctuations are called universal because their magnitude is
independent of the sample parameters such as the mean free path $\ell$, and the
average conductance $\langle g \rangle$. The dependence on sample dimension
is weak. Studies have mainly focused on the
variance of the fluctuations, which is the leading and the universal part of the
fluctuations.

The conductance being a random variable showing such large fluctuations, 
one realized that one should 
consider its full distribution. It was soon clear that the
first higher cumulants of the conductance are proportional to \cite{altshuler3}
\begin{equation}
\langle g^n\rangle_c \propto \langle g \rangle^{2-n} , \qquad n<g_0, \; 
\langle g \rangle \gg1 . 
\label{eqglown}
\end{equation}  
Here the conductance is measured in units of $e^2/\hbar$, $g_0$ is the mean
conductance at the scale $\ell$,  $ \langle \ldots  \rangle $ denotes the
ensemble averaging, and the subscript $c$ indicates cumulants. In
the metallic regime far from localization, where $\langle g \rangle \gg 1$, the higher
cumulants are thus small, and the distribution of the conductance is therefore
roughly Gaussian. However, for $n\agt g_0$ the decrease in 
magnitude of cumulants as described by Eq.~(\ref{eqglown}) is changed 
into a very rapid increase
($\propto \exp[g_0^{-1}n^2]$). This leads to the log-normal
tails of the distribution \cite{altshuler3}. With increasing disorder, the 
log-normal tails become more important. Although the calculation of the full
conductance distribution on the threshold of localization ($\langle g \rangle 
\sim1$) is
today out of reach, it is quite plausible that the whole distribution crosses
over to a log-normal shape in the strongly localized regime (see Refs.
\onlinecite{shapiro2} and \onlinecite{altshulerboek} for a discussion). Indeed,
it is well known that in the strongly localized regime in one dimension
the conductance
is given by the product of transmission amplitudes, yielding a log-normal 
distribution \cite{anderson2,abrikosov2,melnikov,economou2}.

Although high order cumulants govern the tails of the distribution, they do
not affect it near the center. A deviation from the Gaussian distribution near
the center should be revealed, first of all, in the lowest nontrivial
cumulants. An important step in this direction was the recent calculation of
the third cumulant of the distribution using random matrix theory by Macedo
\cite{macedo}. He found for the orthogonal ($\beta=1$) and symplectic ensemble
($\beta=4$) that the third cumulant of the conductance is proportional to
$1/g^2$, thus the leading term given by Eq.~(1) vanishes (see for instance
Ref.\ \onlinecite{metha}  for the definitions of the ensembles).  For the
unitary ensemble ($\beta=2$) even this sub-leading term vanishes. The physical
reason behind this is not clear. (We recently learned that the same result in
quasi one dimension was found by Tartakovski \cite{tartakovskiprive} by the
scaling  method described in Ref.\ \onlinecite{tartakovski}.) However, random
matrix theory is only valid in quasi one dimensional systems. Therefore, it was
not known whether this cancelation is unique to one dimension, or holds also in
higher dimensions, which might indicate an  overlooked symmetry of the system.
That such a possibility exists was also suggested by the fact that the
leading-order contribution to the third cumulant of the density of states
vanishes in two dimensions \cite{altshuler3}.  The question whether
or not the
cancelation holds in two and three dimensions was a major motivation
for this work.

Let us finally mention that related to the conductance distribution, recently
third cumulants \cite{t3pre} and the full distribution functions \cite{probprl}
of two related transmission quantities were calculated. These
quantities, the speckle transmission and the total transmission, were measured
in scattering experiments with electro-magnetic waves (light and microwaves)
\cite{t3prl,genack2}. In the regime considered, these distributions are
independent of dimension. On a diagrammatic level this can be understood: The
diagrams for the cumulants of the speckle and the total transmission are
loop-less diagrams \cite{probprl}. The diagrams for the cumulants of the
conductance, however, contain loops, and will therefore be, a priori, dimension
dependent. For the electro-magnetic waves the Landauer approach was used,
however, for conductance properties this yields the same results as the Kubo
approach \cite{kane,c3pre}. Therefore, the calculations presented 
below are also valid for classical waves.

In the present paper we will calculate the third cumulant of the conductance
for any rectangular geometry in the metallic regime  in quasi one, quasi two,
and three dimensions. In section \ref{secucf} we review the formalism
for calculating conductance fluctuations. In section \ref{secg3} it is
described  how the leading diagrams can be found. In section \ref{seceva} we
calculate the third cumulant and discuss its dimension dependence. After
considering the effects of inelastic scattering, we end with a conclusion.

\section{Conductance fluctuations}\label{secucf}

In this section we introduce our diagrammatic approach, to a large extent
following the detailed paper of Lee, Stone, and Fukuyama \cite{lee2}. We
consider a sample with a rectangular geometry with sizes $L_x \times L_y \times
L_z$; the conductance is measured across the z-direction. The scatterers are
isotropic point scatterers. The scattering is calculated in second order Born
approximation, which assumes weak scattering. (The present calculation
can be easily extended beyond second order Born, this will yield the same
results provided the correct mean free path is taken \cite{thmn4,hik}.) We
consider the metallic, mesoscopic regime, which is characterized by the
inequalities $ 1/k \ll \ell \ll L_z $, where $k$ denotes the Fermi wavenumber;
$\ell$ is the mean free path, related to the scatterer density $n$ and
scatterer strength $v$ as $\ell=4\pi/n|v|^2$. The relaxation time is given by
$\tau=\ell/(2k)$. Incoherent scattering is first neglected, that is, the incoherence
length is much larger than any system size, $L_{\rm in} \gg L_x,L_y, L_z $.

The conductance and its cumulants are calculated with the Kubo formula. The
transport through the sample is given by the so called ladder diagrams or
diffusons, describing the multiple scattering of the electrons on the
scatterers. In the conductance diagram the field creates an electron-hole pair
at a current vertex somewhere in the sample,  after some diffuse propagation
this pair annihilates. This diagram thus contains a bubble, the corresponding
propagator ${\cal L}$  obeys the diffusion equation:
\begin{equation} -\nabla^2_1 {\cal L}({\bf r}_1,{\bf r}_2)=
\frac{1}{2\pi \nu \tau^2 D} \delta({\bf r}_1-{\bf r}_2),\end{equation} 
where $D$ is the diffusion constant. (Correspondence with units often used
with classical waves is found by identifying: $\hbar=1$, $m=1/2$, 
$\nu=k/(4\pi^2)$, $D=2k\ell/3$.)

As mentioned we restrict ourselves to the conductance in the z-direction.
The current is restricted to the z-direction by imposing
fully reflecting boundaries in the
$x$ and $y$ direction, and fully conducting boundaries in the $z$-direction: 
\begin{equation} {\cal L}({\bf r}_1,{\bf r}_2) |_{z_1=0, z_1=L_z}=0 \, , \qquad
\frac{\partial {\cal L}({\bf r}_1,{\bf r}_2)}{\partial x_1} |_{x_1=0, x_1=L_x}
=\frac{\partial {\cal L}({\bf r}_1,{\bf r}_2)}{\partial y_1} |_{y_1=0, y_1=L_y}
=0 ,\end{equation}
 and similarly for the ${\bf r}_2$ dependence.
For rectangular geometries it is useful to write the solution of the diffuson
 propagator in (discrete) momentum space. Following Ref.\onlinecite{lee2}
(in present work there is an extra factor $1/(2\pi \nu \tau)$ in the diffuson), 
we write the diffuson as
\begin{equation} {\cal L}({\bf r}_1,{\bf r}_2)=\frac{1}{2\pi \nu \tau^2 D} 
 \sum_{i_z=1}^\infty \sum_{i_x=0}^\infty \sum_{i_y=0}^\infty
\frac{Q_i({\bf r}_1) Q_i({\bf r}_2)}{\lambda_i} \label{eqLtoQ} \end{equation} 
where $i$ is the momentum vector $(i_x,i_y,i_z) $. 
The $Q_i$'s are the normalized, orthogonal eigenfunctions
\begin{equation} 
Q_i({\bf r})=\sqrt{\frac{2}{L_x}\frac{2}{L_y}\frac{2}{L_z} } \, 
\sin\left(\frac{\pi z i_z}{L_z} \right) \cos\left(\frac{\pi x i_x}{L_x} \right)
 \cos\left(\frac{\pi y i_y}{L_y}\right) . \label{eqdefQ}\end{equation} 
Note that these eigenfunctions are not properly normalized if $i_x=0$ or $i_y=0$, 
in which case
 one has to replace $\cos\left(\frac{\pi x i_x}{L_x} \right)$ 
 or $\cos\left(\frac{\pi y i_y}{L_y} \right)$ by $1/\sqrt{2}$. 
The eigenvalues $\lambda_i$ in Eq.(\ref{eqLtoQ}) are
\begin{equation} \lambda_i= \frac{\pi^2}{L_z^2} \left[ i_z^2 + i_x^2 \left(\frac{L_z}{L_x}
\right)^2 + i_y^2 \left(\frac{L_z}{L_y} \right)^2 \right] \end{equation} 
The longitudinal
momentum $i_z$ takes positive integer values; the transverse momenta $i_x$ and
$i_y$ can in addition also be zero. In quasi two dimensional (2D) geometries
one has $L_x \ll L_z$ and only the  terms with $i_x=0$ contribute in the eigenvalues. 
In quasi one dimensional (1D) samples also $L_y \ll L_z$, and $i_y$ is 
essentially restricted to zero.
On the other hand, for very wide geometries $i_{x}^2 \left( L_z/ {L_{x}}
\right)^2$ becomes a continuous variable and  the sum over the
$x$-variable (or $y$-variable) becomes an integral. The average value of the
dimension-less conductance $\langle g \rangle $ is
\begin{equation} \langle g \rangle = 2\pi \nu D \frac{L_x L_y}{L_z}.
 \end{equation} 
The inverse of this, $1/\langle g \rangle$, will be a small parameter 
in our diagrammatic expansion.

\subsection{Hikami-boxes}

\begin{figure}
\epsfxsize=9.5cm
\centerline{\epsffile{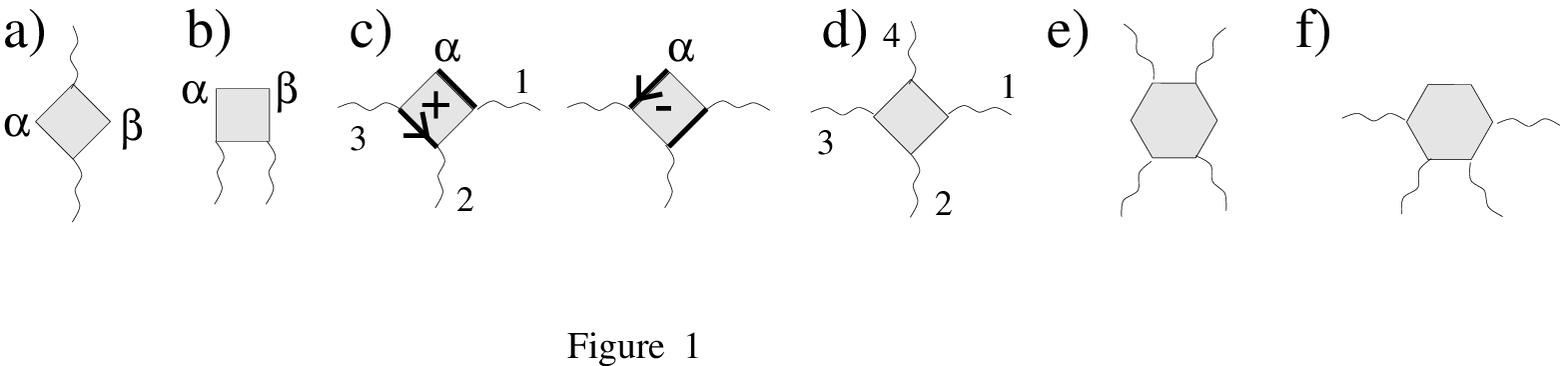}}
\caption{Hikami four-point and six-point vertices used. 
The wavy lines denote diffusons or cooperons. The 
current vertices are at the unoccupied corners of the polygons and are 
labeled by $\alpha$ and $\beta$.
\label{figboxes}}
\end{figure}

For the calculation of the second and higher order cumulants of $g$, we need
the vertices describing the interference between two diffusons. These vertices
are known as Hikami-boxes\cite{hikami,gorkov}. Formally they arise from the
spatial derivatives $\partial / \partial {\bf r}_{\alpha}$ 
and $\partial / \partial {\bf r'}_{\beta}$
in the Kubo formula for $\sigma_{\alpha,\beta}({\bf r}, {\bf r'})$, 
where the indices $\alpha, \beta=x,y,z$ label the directions of the incoming and
outgoing current. We have drawn the boxes in Fig.\
\ref{figboxes}, the corresponding expressions are labeled $H_a$ to $H_f$,
respectively: 
\begin{mathletters}
\begin{eqnarray} 
H_a&=&4\pi \nu \tau^3 \frac{\delta_{\alpha,\beta}}{3}, \label{eqhika} \\
H_b&=&2\pi \nu \tau^3 \frac{\delta_{\alpha,\beta}}{3}, \label{eqhikb}\\ 
H_{c1}&=&  4 \pi i \nu \tau^4 {\bf q}^{\alpha}_2  \frac{k}{3}, \\
H_{c2}&=&  -H_{c1}, \\
H_d&=&2\pi\nu \tau^4 D \left[ -2 {\bf q}_1 \cdot {\bf q}_3
-2 {\bf q}_2 \cdot {\bf q}_4-({\bf q}_1+ {\bf q}_3)\cdot ({\bf q}_2+ {\bf
q}_4) \right] , \\ 
H_e&=&4\pi \nu \tau^3 \frac{\delta_{\alpha,\beta}}{3}, \\
H_f&=&-2\pi \nu \tau^5 \frac{\delta_{\alpha,\beta}}{3}.
\end{eqnarray} 
\end{mathletters}
The ${\bf q}$'s are the momenta of the
diffusons (the wavy lines), as numbered according Fig.\ \ref{figboxes}. 
As the conductance is calculated in the z-direction, $\alpha$ and
$\beta$ are taken $z$. Likewise, only the z-component of the momentum in $H_c$
will be taken into account. $H_c$ comes in two flavors: $H_{c1}$ and $H_{c2}$. 
$H_{c2}$ has an additional minus sign due to the reversed 
orientation of the advanced and retarded propagators of the current vertex. 
We used that the diffusons vary slowly on length scales of the mean free path, 
i.e. $ q\ell \ll 1$. Finally, other boxes are sub-leading.

\subsection{UCF calculation}

\begin{figure}
\epsfxsize=2.5cm
\centerline{\epsffile{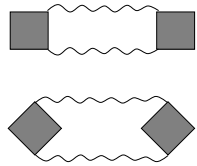}}
\caption{The UCF or $\langle g^2 \rangle_c$ diagrams.
 \label{figg2kubo}}
\end{figure}

For clarity we briefly present the UCF calculation using this formalism.  The
UCF-diagrams are connected diagrams containing two conductivity bubbles and
thus four current vertices.  Its diagrams are  shown in Fig.\ \ref{figg2kubo}.
The UCF diagrams contain two four-boxes ($H_a$ or $H_b$) and two diffusons. 
For the orthogonal ensemble the combinatorial factor for the upper diagram is
4. The lower diagram has combinatorial factor 2,  and the two Hikami vertices
bring, comparing Eq.\ (\ref{eqhika}) to Eq.\ (\ref{eqhikb}), 
an additional factor of $2^2$. Thus one
finds a pre-factor 12 for the sum of all diagrams with respect to a single
diagram of the upper topology.  (In the unitary ensemble cooperons do not
contribute and this would be a factor 6.)
For each incoming or outgoing current vertex there is a factor $2k/L_z$. It
arises from the expression $1/(mL_z) \to 2/L_z$, that can be read  off from
Eq.(2.5) in Ref. \onlinecite{lee2}, and the wavenumber $k$ that originates 
from each derivative in the Kubo formula.
Next, we use the Fourier decomposition of the diffusons Eq.(\ref{eqLtoQ})
and interchange the sum over the eigenvalues with the spatial
integrals. The two diffusons interfering at the Hikami-box yield
the orthogonality relations: $\int d{\bf r} Q_i({\bf r}) Q_j({\bf r}) =
\delta^{(3)}_{i,j}$. One finds
\begin{eqnarray} \langle g^2 \rangle_c &=& 12\left(\frac{2k}{L_z}\right)^4 
\left(\frac{1}{2\pi \nu \tau^2 D}
\right)^2 \left(\frac{2\pi\nu \tau^3}{3}\right)^2 \int
d{\bf r}_1
\int d{\bf r}_2 {\cal L}({\bf r}_1,{\bf r}_2)^2 \nonumber \\ 
&=&\frac{12}{L_z^4} \sum_{i,j}\frac{1}{\lambda_i \lambda_j}
\int d{\bf r}_1 Q_i({\bf r}_1) Q_j({\bf r}_1) 
\int d{\bf r}_2 Q_i({\bf r}_2)  Q_j({\bf r}_2) \nonumber \\
&=&\frac{12}{L_z^4} \sum_{i,j}\frac{1}{\lambda_i \lambda_j}
(\delta_{i_z,j_z} \delta_{i_x,j_x }\delta_{i_y,j_y})^2 \nonumber \\
&=&\frac{12}{\pi^4} \sum_{i_z=1,i_x=0,i_y=0}^\infty
 \left(i_z^2+i_x \frac{L_z^2}{L_x^2} +i_y\frac{L_z^2}{L_y^2}\right)^{-4} .
\end{eqnarray}
In quasi one dimension this yields $\frac{12}{\pi^4} \sum_{i_z=1}^{\infty} 
i_z^{-4} =\frac{2}{15}$, 
which is the well-known result for the UCF in quasi one dimension\cite{lee2}.

\section{The third cumulant}\label{secg3}

\begin{figure}
\epsfxsize=5.5cm
\centerline{\epsffile{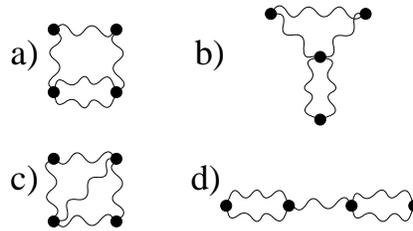}}
\caption{All basic topologies for the third cumulant diagrams. 
The wavy lines are diffusons or cooperons. 
The dots represent Hikami four point vertices. 
Detailed inspection of these diagrams will yield the diagrams of 
Fig.\ \protect{\ref{figtop}}.
\label{figbasictop}}
\end{figure}

We now study the third cumulant.
First, we rewrite Eq.(\ref{eqglown}) into a relation for the relative cumulants: $
\frac{\langle g^n \rangle_c}{\langle g \rangle^n} \propto \langle
g\rangle^{2-2n}$. The inverse power of $\langle g \rangle $ on the right
hand side can be interpreted as the number of Hikami four-boxes in the
diagrams.  Thus the third cumulant diagram contains four Hikami four-boxes.
Indeed, it proves impossible to create $\langle g^3 \rangle_c$ diagrams with
less boxes. Diagrams with more boxes are sub-leading as they are of
higher power of $1/ \langle g \rangle$, which is a small parameter.
In diagrammatic approaches it is often tricky to find all the
relevant diagrams; already for the much simpler set of UCF diagrams there was
considerable discussion in the literature. We used the following considerations
to find the set of diagrams for the third cumulant: The diagrams have six
current vertices: three incoming ones and three outgoing ones. The diagrams
contain four 4-point vertices and  each vertex has at
least two diffusons attached to it, see Fig.\ \ref{figboxes}. This leaves two
possibilities: there are three boxes with each two diffusons and one box with
four diffusons; or, are two boxes with two diffusons and two boxes with
three diffusons each. It is now an easy exercise to see that there are only
four possible basic topologies. They are drawn in Fig.\ \ref{figbasictop}. The
dots represent the 4-point vertices; the lines represent diffusons. The next step is to
insert the Hikami-boxes and connect them in all possible ways to the diffusons.
This yields many diagrams, yet not in all diagrams the outgoing electron-hole
pair is the same as the incoming electron-hole pairing. Such diagrams are not
the product of three conductance bubbles; they do not contribute to the
$\langle g^3 \rangle_c $ process and are left out. We end up with the diagrams
represented in Fig.\ \ref{figtop}. In the figure there are also diagrams of the
same order  with one six-box and two four-boxes. They can be obtained by
contracting one diffuson in the diagrams with four-boxes. In physical terms the
six-box diagrams correspond to processes where after an interference process,
the amplitudes do not combine into a diffuson.  Instead, the amplitudes
interact again directly without being scattered.

\subsection{Evaluation of the diagrams}\label{seceva} 

\begin{figure} 
\epsfxsize=12cm 
\centerline{\epsffile{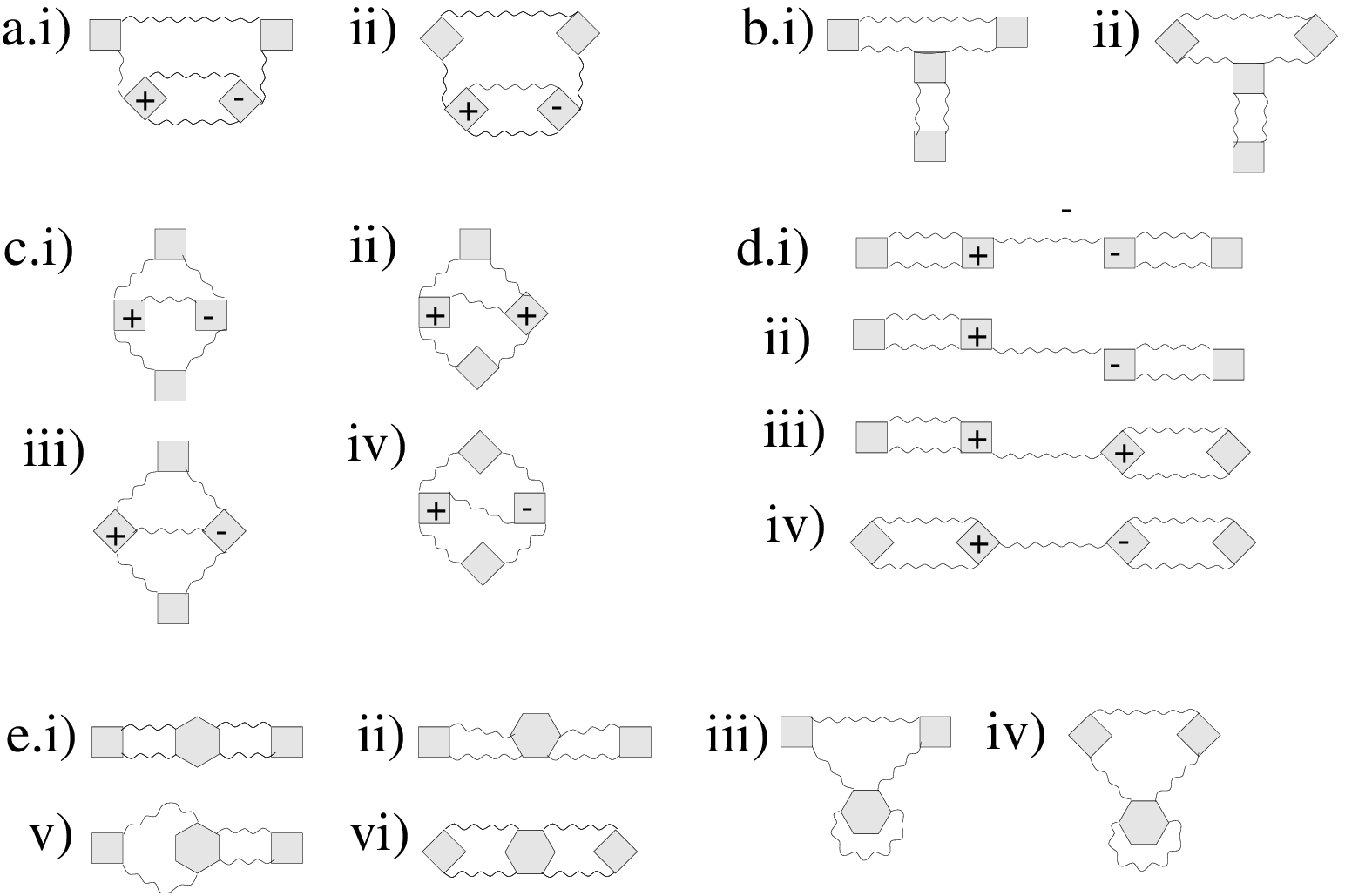}}
\caption{The calculated set of leading diagrams for the third cumulant of 
the conductance. The diagrams are derived from Fig.\
\protect{\ref{figbasictop}}; one sees that structure is the similar.
The diagrams e) are obtained by contraction
of diffusons in the diagrams a), b), c) and d). \label{figtop}} \end{figure}

We calculate the diagrams for quasi-1D, quasi-2D, and 3D geometries. The
calculation goes similar to the UCF-calculation, that is, the sums over the
momenta are taken outside the spatial integrals. If two internal
diffusons interfere, the spatial integrals bring the orthogonality relations as
used for the UCF calculation. However, two complications arise. First, at
the Hikami vertices not only two, also three, or four diffusons 
interfere (as is seen from from the Figs.\ \ref{figbasictop} and \ref{figtop}),
this gives more complicated structures. Secondly, Hikami-boxes $H_c$ and $H_d$
are $q$-dependent, corresponding to spatial derivatives of the
diffusons. We therefore introduce the matrices $R$, $S$, and $T$, which 
describe the interaction of the 
diffusons at the vertices.
                     
In the diagrams e.i), e.ii), e.v), and e.vi) of Fig.\ \ref{figtop} four 
diffusons, 
two with momentum $i$, and two with momentum $j$ interfere. The x-components 
of the diffusons couple as 
\begin{equation} R_{i_x,j_x}= \left( \frac{2}{L_x} \right)^2 \int_0^{L_x}
dx \cos^2 \left( \frac{i_x x \pi}{L_x} \right) 
\cos^2\left( \frac{j_x x \pi}{L_x} \right)= 
\frac{1}{L_x} \left[ 1+ \frac{1}{2} (1-\delta_{i_x,0}) \delta_{i_x,j_x}
\right] \end{equation} 
The y-components yield exactly the same integral, see Eq.(\ref{eqdefQ}), and
also the z-components yield the same expression. These diagrams will thus be
proportional to $R_{i_z,j_z} R_{i_x,j_x} R_{i_y,j_y}$.

In most other diagrams three diffusons interfere at a
Hikami box $H_{c1}$, this box brings a derivative $\partial/ \partial_z$.
The matrix $S$ describes the effect of this derivative
on the z-component, the $T$ describes the integrals over x and y components.
\begin{eqnarray}
S_{i_z,j_z,k_z}&=&\left(\frac{2}{L_z} \right)^{3/2} \int_0^{L_z} dz
\left( \frac{i_z\pi}{L_z} \right) \cos\left(\frac{i_z z \pi}{L_z} \right)
\sin\left(\frac{j_z z \pi}{L_z} \right)
\sin\left(\frac{k_z z \pi}{L_z} \right) \nonumber \\
&=&\frac{i_z\pi}{L_z\sqrt{2L_z}}
  ( -\delta_{i_z,j_z+k_z}+\delta_{j_z,i_z+k_z}+\delta_{k_z,i_z+j_z} )\\
T_{i_x,j_x,k_x}&=& \left(\frac{2}{L_x} \right)^{3/2} \int_0^{L_x} dx 
\cos\left(\frac{i_z z \pi}{L_z} \right) \cos\left(\frac{j_z z \pi}{L_z} \right)
\cos\left(\frac{k_z z \pi}{L_z} \right) \nonumber \\
&=& \frac{1}{\sqrt{L_x}} (\delta_{i_x,0}\delta_{j_x,k_x}  \!+\!
\delta_{j_x,0}\delta_{i_x,k_x} \! + \!
 \delta_{k_x,0}\delta_{i_x,j_x} \!-\!2 \delta_{i_x,0} 
\delta_{j_x,0}\delta_{k_x,0} )
\quad  \mbox{(if $i_x$, $j_x$, or $k_x=0$)} \\
 &=& \frac{1}{\sqrt{2L_x} } (\delta_{i_x,j_x+k_x}+
 \delta_{j_x,i_x+k_x}+\delta_{k_x,i_x+j_x}) 
 \qquad \qquad \mbox{(else)} \end{eqnarray}
The box $H_{c2}$ has a minus sign resulting in 
$-S_{i_z,j_z,k_z}T_{i_x,j_x,k_x}T_{i_y,j_y,k_y}$.

The diagrams can now be written in terms of the $R, S, T$. 
The pre-factors, the combinatorial factors, and the sum over the
momenta are included. The third cumulant is given by the sum of diagrams
$\langle g^3 \rangle_c = \sum F$, where the $F$'s are,
\begin{mathletters}
\begin{eqnarray}
F_{a.i}+F_{a.ii}&=& -144 r \sum_{i,j,k} \frac{1}{\lambda_i^3 \lambda_j \lambda_k} \,
 S_{i_z,j_z,k_z}^2 T_{i_x,j_x,k_x}^2 T_{i_y,j_y,k_y}^2 \label{eqFai} \\
F_{b.i}+F_{b.ii}+F_{e.iii}+F_{e.iv} &=& 72 r \sum_{i,j} \frac{1}{\lambda_i^3 \lambda_j^2} 
 \, \left[\frac{\pi^2}{L_z^3} \left( \frac{3}{2} i_z^2 \delta_{i_z,j_z}
\! +\!i_z^2 \right)
R_{i_x,j_x} R_{i_y,j_y} \right. \nonumber \\ && 
\left. +\frac{\pi^2}{L_x^3} \left( 
\frac{3}{2} i_x^2 \delta_{i_x,j_x} \!+ \!i_x^2 \right)
R_{i_z,j_z} R_{i_y,j_y}+\frac{\pi^2}{L_y^3} \left(
 \frac{3}{2} i_y^2 \delta_{i_y,j_y} \!+\!i_y^2 \right) R_{i_x,j_x} R_{i_z,j_z}
\right]  \label{eqFb} \\
F_{c.i}&=& -48 r \sum_{i,j,k} \frac{1}{\lambda_i^2 
\lambda_j^2 \lambda_k} \,
S_{i_z,j_z,k_z}^2 T_{i_x,j_x,k_x}^2 T_{i_y,j_y,k_y}^2 \\
F_{c.ii}+F_{c.iii}
&=&-120 r \sum_{i,j,k} \frac{1}{\lambda_i^2 \lambda_j^2 \lambda_k} \,
S_{k_z,i_z,j_z}^2 T_{i_x,j_x,k_x}^2 T_{i_y,j_y,k_y}^2  \\ 
F_{c.iv}&=& 96 r\sum_{i,j,k} \frac{1}{\lambda_i^2 \lambda_j^2 \lambda_k} \,
 S_{i_z,j_z,k_z} S_{j_z,i_z,k_z} T_{i_x,j_x,k_x}^2 T_{i_y,j_y,k_y}^2  \\ 
F_{d.i}+F_{d.ii}+F_{d.iii}+F_{d.iv}
&=&864 r \sum_{i,j,k} \frac{1}{\lambda_i^2 \lambda_j^2 \lambda_k } \,
S_{i_z,i_z,k_z} S_{j_z,j_z,k_z} T_{i_x,i_x,k_x}T_{j_x,j_x,k_x}
T_{i_y,i_y,k_y}T_{j_y,j_y,k_y} \\
F_{e.i}+F_{e.ii}+F_{e.v}+F_{e.vi}
&=&36r \sum_{i,j} \frac{1}{\lambda_i^2 \lambda_j^2} \, R_{i_z,j_z}
R_{i_x,j_x} R_{i_y,j_y}.
\end{eqnarray} 
\end{mathletters}
The sums runs over all allowed momenta indices. The factor $r$ contains the
pre-factors of the diffusons and boxes and a factor $(2k/L_z)^6$ for the
incoming and outgoing current vertices, \begin{equation} 
r = \frac{1}{2\pi \nu D L_z^6} \end{equation} 
Finally, the sums still contain a factor $L_z^7 /(L_x L_y)$, yielding
multiplied by $r$  a pre-factor $\langle g\rangle ^{-1}$. So indeed  the third
cumulant is proportional to the inverse dimension-less conductance as predicted.
The choice of ensemble is reflected in the combinatorial factor. For
simplicity the combinatorial factors were calculated for the unitary ensemble,
finally, for the orthogonal ensemble all pre-factors will be four times larger. 
The e.iii)
and e.iv) diagrams are divergent in two and three dimensions, as the sum
$\sum_{j_z,j_x,j_y} \lambda_j^{-1}$ is logarithmically divergent in 2D and
linearly divergent in 3D. This divergence is, however, exactly canceled by a
similar divergence in the b.i) and b.ii) diagrams. Eq.~(\ref{eqFb}) presents the
combined, finite expression.

It is
interesting that due to the finite system size momentum conservation apparently
does not hold, for example,  the momentum of the middle diffuson in the d)
diagrams of Fig.\ \ref{figtop} need not be zero. This is a result of the
mirror terms in the diffusons, which are present to fulfill to the 
boundary conditions. Though such terms would be absent in the bulk, they never
vanish in our geometry.

In the quasi one-dimensional case the diffusons are simple linear functions of
the $z-$coordinate  
\begin{equation}
{\cal L}(z_1,z_2) = \frac{1}{2\pi \nu \tau^2 D} \frac{{\rm min}(z_1,z_2)
 [L_z-{\rm max}(z_1,z_2)]}{L_z} , \end{equation}
 allowing an analytically treatment.
The diagrams can now even be
calculated directly without introducing the momentum representation at all. As
a check we also calculated the diagrams using this representation. We find,
using either method, for the sum of the diagrams \begin{equation} \langle g^3
\rangle_c =0 \qquad \mbox{(quasi 1D).} \end{equation}  Thus the leading
contribution to the third cumulant in one dimension vanishes. This confirms the
random matrix theory result \cite{macedo} diagrammatically.

\begin{figure}
\epsfxsize=8cm
\centerline{\epsffile{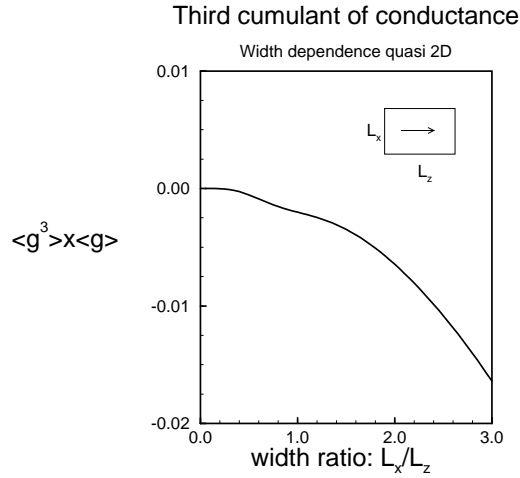}}
\caption{The third cumulant of the conductance 
multiplied by the average as a function of the transversal size in a 2D
sample. Although the leading term 
vanishes in quasi 1D, it is non-trivial in 2D geometries. \label{figwide2}}
\end{figure}

\begin{figure}
\epsfxsize=8cm
\centerline{\epsffile{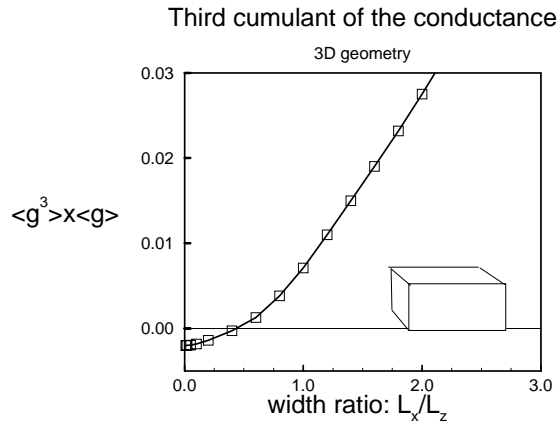}}
\caption{The third cumulant multiplied by the average conductance
in 3D sample plotted against the transversal size $L_x$. 
The geometry is $L_x \times L_z \times L_z$ for the solid line.
Note that, the sign changes in going from a quasi-2D to a 3D sample.
The dashed line gives third cumulant in sample with 
geometry $L_x \times L_x \times L_z$  \label{figwide3}.}
\end{figure}

In higher dimensions the sums were performed
numerically. In the numerical evaluation the sums over three sets of momenta
are implicitly reduced to sums over two sets. This is a direct consequence of
the fact that there are only two independent momenta present. The numerical
evaluation remains, however, quite involved as in three dimensions one still
has to sum over six variables and the convergence is quite slow. We find in
the orthogonal ensemble for square, and cubic samples, resp. \begin{eqnarray}
\langle g^3
\rangle_c&=& -0.0020  \langle g \rangle^{-1} \qquad \mbox{ (quasi
2D)} \nonumber \\ \langle g^3 \rangle_c&=& + 0.0076  \langle g
\rangle^{-1} \qquad \mbox{ (3D)} \end{eqnarray} 
The results for rectangular samples are
given in the figures \ref{figwide2} and \ref{figwide3}, where we multiplied
the third cumulant by the average of the dimension-less conductance. The
third cumulant for wide slabs ($L_x \gg L_z , \, L_y \gg L_z$) is proportional
to $(L_x L_y / \langle g \rangle)$. In the figures due to the multiplication by 
$\langle g \rangle$ there is 
proportionality to $(L_x L_y)^2/L_z^4$ for wide slabs. 
For very narrow slabs one sees that the correct quasi 2D and quasi
1D limits are recovered. Note that third cumulant passes through zero when
going from 2D to 3D, this happens if the sample has the size $ 0.46 L_z \times
L_z \times L_z $.  

\subsection{Inelastic scattering}

Incoherent scattering was neglected in above calculations. In
realistic system, however, incoherent scattering can be present. 
This is especially true at non-zero temperatures where electron-phonon
interactions will occur.
This mechanism was included in the description of Lee, Stone, and Fukuyama
\cite{lee2}. The inverse inelastic scattering time induces 
a positive shift of the diffuson eigenvalues 
\begin{equation}
\lambda_i=\frac{\pi^2}{L_z^2} \left[
i_z^2+i_x^2 \left( \frac{L_z}{L_x} \right)+ 
i_y^2 \left( \frac{L_z}{L_y} \right) 
+ \frac{L_z^2 }{\pi^2  L_{\rm in} } \right]
\end{equation}

The case that the incoherence length is much smaller than the system sizes is of
particular interest. The effect on the third cumulant can now be estimated
simply by
considering the sample as being made up of small samples of dimension 
$L_{\rm in}^d$ with independent conductance distributions. 
The conductance of such a sample is in 3D
\begin{eqnarray}
g&=& \sum_{i_x}^{L_x/L_{\rm in} }  \sum_{i_y}^{L_y/L_{\rm in} } 
\left[ \sum_{i_z}^{L_z/L_{\rm in}} \left( g_0 \right)^{-1} \right]^{-1}
\nonumber \\
&=& \frac{L_x L_y}{L_z L_{\rm in} }  g_0+
\frac{L_{\rm in}^2}{L_x L_y} \sum_i
\delta g_0,
\end{eqnarray}
where $g_0$ denotes the conductance of a individual coherent cube.
The relative cumulants thus scale as
\begin{equation}
\frac{\langle g^n \rangle_c} {\langle g \rangle^n} = 
\left( \frac{L_{\rm in}^3}{L_x L_y L_z} \right)^n 
\frac{\langle g_0^n \rangle_c} {\langle g_0 \rangle^n} .
\end{equation}
The relative third cumulant reduces by a factor $L_{\rm in}^6/(L_x
L_y L_z)^2$. As a result the combination
\begin{equation}
\frac{\langle g^3 \rangle_c \langle g \rangle}{\langle g^2 \rangle_c^2}
\end{equation}
is independent of $L_{\rm in}$, so that this quantity is universal.  As the
incoherence length is the same in all directions, the coherent parts are
essentially cubic. Therefore, the quantity will roughly tend to its value for
{\em coherent, cubic} samples.  The precise value can be obtained by properly
including the incoherence effects in the calculation as indicated above.
Because of this universality, the quantity is useful experimentally. We even
expect the full conductance distribution to be universal in this regime, that
is, independent of geometry and incoherence length.

\section{Conclusion} 

We have considered a mesoscopic sample in the metallic regime and we calculated
the third cumulant of the conductance distribution. Naive scaling predicts that
the third cumulant should be proportional to $1/\langle g \rangle $. In quasi
one dimension we confirm the absence of this leading contribution, as was
found by Macedo\cite{macedo}. In two and three dimensions, however,
this is cancelation is not present; 
the leading contribution to the third cumulant is negative in
two dimensions and positive in three dimensions. The fact that the third
cumulant changes sign is surprising. The third cumulant is also known as the
skewness of a distribution. In analogy with the third cumulant of the total
transmission \cite{probprl} or if the distribution would be tending to
log-normal, one would have expected a positive third cumulant of the
conductance. Instead, we find that all possible values occur: negative,
positive and zero. We have no explanation for this. 

To the best of our
knowledge there exists no experimental work where the conductance distribution
is discussed. Experiments, either electronical, or using classical waves, or
numerical simulations, could enlighten present results. Predicted values are
small, but should detectable in electronic systems with moderate values of
$\langle g \rangle$. Also observation of the conductivity distribution as a
whole would be very interesting.

\acknowledgements{
Th.M.N.\ thanks M. Sanquer for discussion. 
Two of us (I.V.L.\ and B.L.A.) gratefully acknowledge support of the NSF under 
Grant No.\ PHY94-07194 and kind hospitality extended to us at ITP in 
Santa Barbara at the final stage of this work.
This research was also supported by N.A.T.O.\ (grant nr.\ CRG 921399).}

\end{document}